# Structure-Dynamics Correlation in Metallic Glass Revealed by 5-Dimensional Scanning Transmission Electron Microscopy

Katsuaki Nakazawa, Kazutaka Mitsuishi, Iakoubovskii Konstantin, Shinji Kohara, and Koichi Tsuchiya

National Institute for Materials Science, 1-2-1, Sengen, Tsukuba, Ibaraki 305-0047, Japan

**Abstract**

Dynamic and structural heterogeneities play an important role in glass transition phenomena and in the formation of amorphous structures. Since structure and dynamics are mutually related, it is expected that there exists some relation between them; however, this relation has not been characterized by a direct experiment. Elucidation of this relation is the key to identifying the structure responsible for the rapid freezing of atomic motion during the glass transition. In this study, we simultaneously observed the dynamic and structural heterogeneities near the glass transition temperature in $Zr_{50}Cu_{40}Al_{10}$ using five-dimensional scanning transmission electron microscopy, which is capable of recording the spatiotemporal distribution of electron diffraction pattern. Dynamic and structural heterogeneities were visualized with sub-nanometer resolution upon heating *in situ,* and a spatial correlation between them was observed up to the glass transition temperature. Simultaneous measurements of dynamic and structural heterogeneities directly revealed that the ordered atomic structure had slow dynamics and that the order decreased with temperature.

**Introduction**

Glass transition is a universal phenomenon observed in a wide range of materials, including metals, polymers, and ceramics, which were cooled from the melt quickly enough to avoid crystallization and freeze the constituent atoms or molecules in disordered positions[1,2]. Since the glass phase has different physical properties compared with a liquid or a crystal, and the glass transition is strongly related to crystallization and melting, this phenomenon is important for controlling physical and technological properties such as moldability[3]. However, the process of freezing the atomic dynamics is still not understood. In the glass transition phenomenon, the viscosity changes rapidly near the glass transition temperature ($T_g$), even though the *average* structure remains almost unchanged[4]. Hence, in an attempt to explain the viscosity variations, researchers focused on the *microscopic* structure of glasses and supercooled liquids. Dynamic heterogeneity was discovered in simulations and experiments on colloidal systems[5,6], in which the motion of particles is heterogeneous and results in clustering. The dynamic heterogeneity shows a divergent behavior near $T_g$ and is expected to explain the drastic change in viscosity during the glass transition. Recent studies using dark-field electron correlation microscopy have directly visualized the dynamic heterogeneity[7,8]. Furthermore, it has been revealed by electron tomography and other methods that there is a spatial heterogeneity of atomic arrangement in glasses[9–13]. This heterogeneity is called structural heterogeneity.

The relation between dynamic and structural heterogeneities is important. If a common atomic structure of the slow-dynamics region in the dynamic heterogeneity is detected, it may become possible to identify the structures that cause the slow dynamics (high viscosity) and are necessary for the glass transition. Since atomic motion is affected by the surrounding atomic structure, there must be a relation between them. Such a relation has been suggested,[8,14–16] but has not been confirmed experimentally so far, possibly because of the scarcity of observation methods that can simultaneously probe both the structural and dynamic heterogeneities. Those methods require a spatial resolution of a few nanometers, a time resolution of a few seconds, and the ability to measure the local structure of glass. A novel technique of 5-dimensional scanning transmission electron microscopy (5D-STEM)[17–19] that involves convergent electron beam diffraction (CBED) does qualify these requirements. 5D-STEM can measure the spatiotemporal distribution of CBED patterns, and these patterns can yield local information that is hard to retrieve in the conventional parallel-beam diffraction mode. When an electron beam irradiates a macroscale volume of a glass, a halo pattern is observed. However, when the probed volume shrinks to the nanoscale, intensity fluctuations, called speckles, appear instead[20,21]. The speckle pattern reflects the local order of the atomic structure[22–25] and its temporal change reflects the motion of the local atomic structure[26]. Thus, as illustrated in Fig. 1, the local structure of a glass can be analyzed from the spatial distribution of diffraction patterns, and the local atomic rearrangement can be measured from the corresponding time sequences.

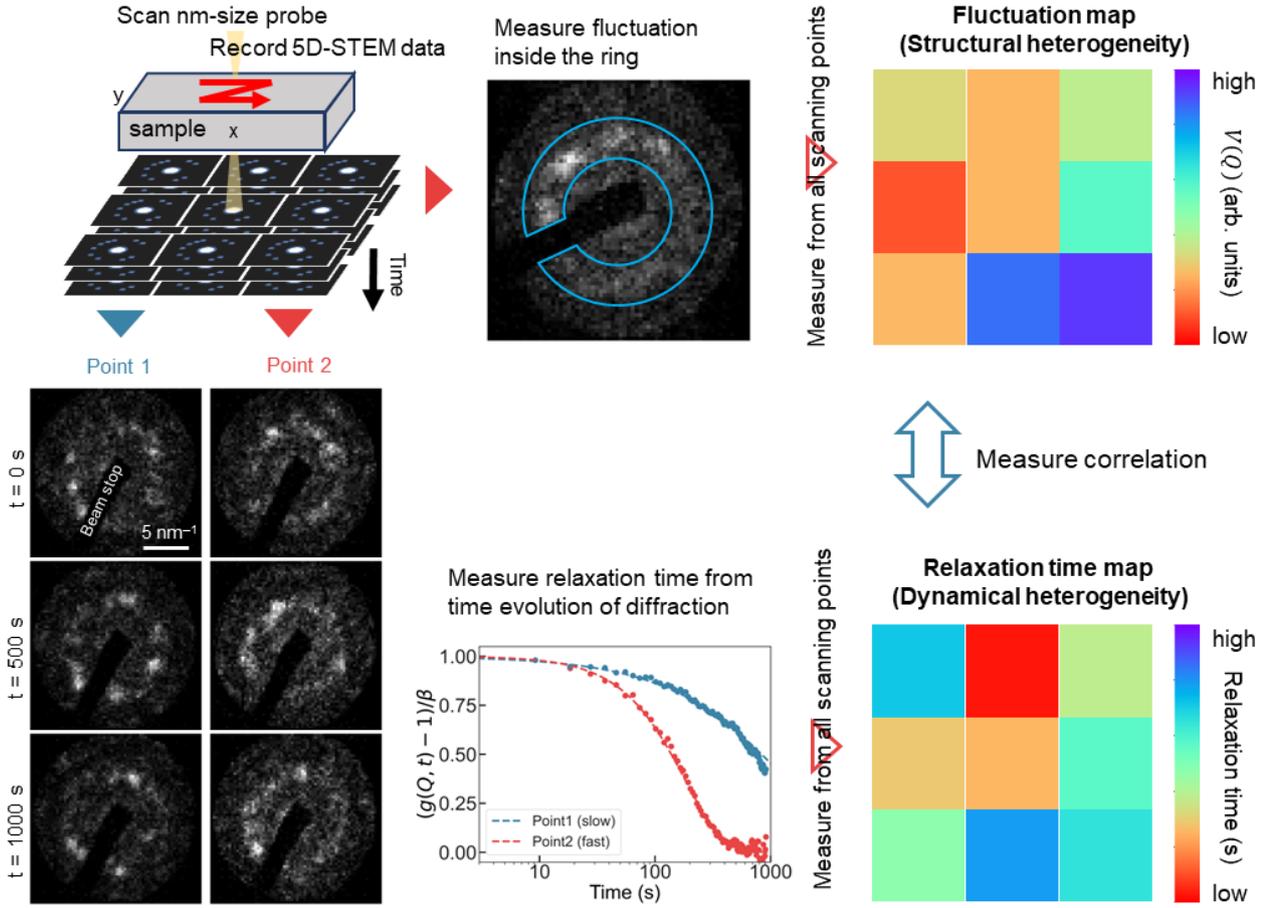

**Fig. 1**. Schematic of measurements of dynamical and structural heterogeneities by 5D-STEM.

In this study, we used 5D-STEM to simultaneously observe dynamic and structural heterogeneities and analyze the relation between them in a $Zr_{50}Cu_{40}Al_{10}$ metallic glass. We also conducted an *in situ* heating experiment to measure the temperature dependences of the dynamic and structural heterogeneities and the structure-dynamics relation below $T_g$.

**Results and Discussion**

5D-STEM measurements were performed in a $Zr_{50}Cu_{40}Al_{10}$ metallic glass with a $T_g$ of 673 K[27]. First, we heated the sample to 633 K. To investigate temperature dependences of dynamic and structural heterogeneities and structure-dynamics relations, we heated the sample from 633 K to 673 K in 10 K increments. The local dynamics was evaluated from the temporal change of the local diffraction patterns using the following equations:[28]

$$G(Q, t_1, t_2) = \frac{\langle I(Q, t_1)I(Q, t_2)\rangle_k}{\langle I(Q, t_1)\rangle_k \langle I(Q, t_2)\rangle_k}.$$

$$g(Q, t) = \langle G(Q, t_1, t)\rangle_{t_1}.$$

Here $Q$ is the spatial frequency and $t$, $t_1$ and $t_2$ are acquisition times. $G(Q, t_1, t_2)$ measures the correlation between diffractions acquired at times $t_1$ and $t_2$. $I(Q, t_1)$ is the signal intensity, and $\langle\ \rangle_k$ indicates averaging over spatial frequency. In this study, we limited the frequency to the range 4.2-4.6 nm$^{-1}$ that covered nearly all speckles. Speckles are formed when the electron beam irradiates an ordered atomic structure along its symmetry axis. Thus, this method probes the dynamics of the ordered atomic structures whose symmetry axis is aligned to the beam direction. $g(Q, t)$ takes the average of $G(Q, t_1, t_2)$ over $t_1$, at a fixed delay time, $t = t_2 - t_1$. The Kohlrausch-Williams-Watts (KWW) function was fitted to $g(Q, t)$ to measure the dynamics.

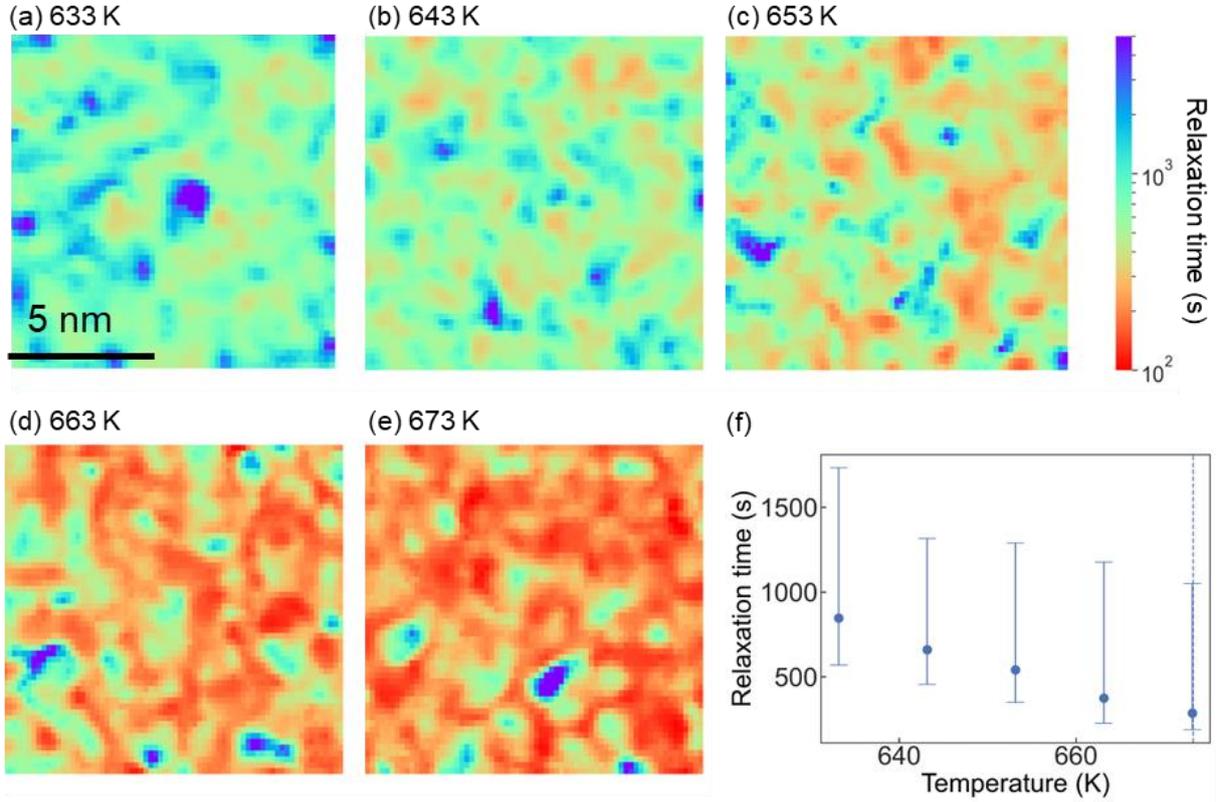

**Fig. 2**. Maps of relaxation time at (a) 633 K, (b) 643 K, (c) 653 K, (d) 663 K, and (e) 673 K. (f) Temperature dependence of the spatially averaged relaxation time.

$$g(Q, t) - 1 = \beta \exp\left(-2\left(\frac{t}{\tau}\right)^\gamma\right)$$

Here $\beta$, $\tau$ and $\gamma$ are a scaling constant, the relaxation time and the stretch factor, respectively. The stretch factor represents the nonlinearity of relaxation and indicates whether relaxation is compressed ($\gamma > 1$) or stretched ($\gamma < 1$) as compared to the simple exponent. Relaxation time is an indicator of dynamics; long relaxation time indicates slow dynamics and *vice versa*. We measured the relaxation time from all scan points to obtain a relaxation time map that directly reflects the dynamical heterogeneity. The relaxation time map for 633 K is presented in Fig. 2(a). As seen from the map, the relaxation time spatially varies from 200 s to 5,000 s revealing a heterogeneity of local dynamics. As shown in Figs. 2(b)-(e), although the relaxation times vary, the dynamic heterogeneity has been observed at different temperatures. The average relaxation time is plotted in Fig. 2(f); it shortens as the temperature approaches $T_g$, indicating an increase in atomic mobility. In samples with similar composition ($Cu_{59}Zr_{41}$), the relaxation process below $T_g$ is thought of as $\beta$-relaxation. From the energy landscape perspective, $\beta$-relaxation is recognized as hopping across sub-basins inside an identical mega-basin. The activation energy to cross the sub-basins can be estimated from the temperature dependence of the relaxation time below $T_g$. The relation between relaxation time and activation energy is expressed by the Arrhenius law.

$$\tau = \tau \exp\left(\frac{E_\beta}{RT}\right),$$

where $E_\beta$ is the activation energy and $R$ is the gas constant. The activation energy was calculated as 97 kJ/mol. This value is similar to the previously reported value of 46 kJ/mol deduced for $Cu_{59}Zr_{41}$ from internal friction

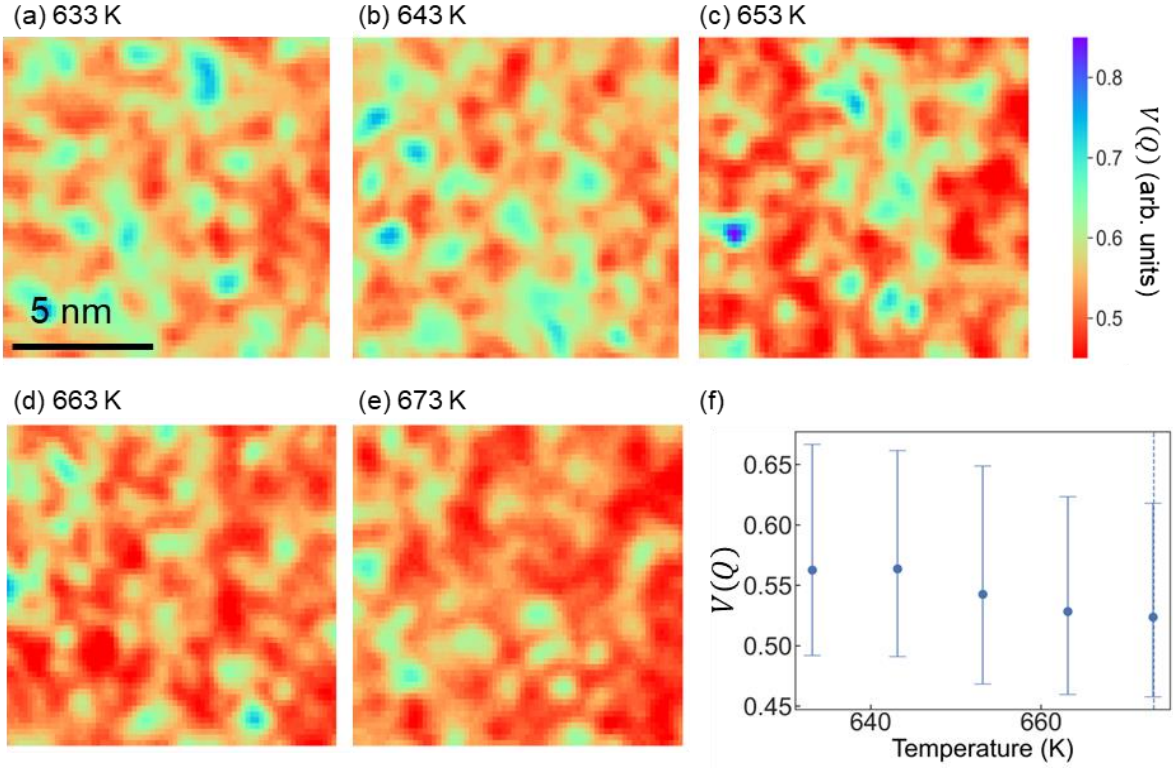

**Fig. 3**. Maps of $V(Q)$ at (a)633 K, (b)643 K, (c)653 K, (d)663 K, and (e)673 K. (f)The average $V(Q)$ at all temperatures. The errorbar shows the distribution of $V(Q)$.dependence of the spatially averaged relaxation time.

measurements[29].

To evaluate the structural heterogeneity, we measured the intensity fluctuation along the azimuthal direction around the halo pattern. This intensity fluctuation mainly originates from the speckle pattern. The fluctuation is measured by the following equation:

$$V(Q, \mathbf{r}) = \frac{\langle I(\mathbf{Q}, \mathbf{r})^2 \rangle_\mathbf{Q}}{\langle I(\mathbf{Q}, \mathbf{r}) \rangle_\mathbf{Q}^2} - 1.$$

Since the speckle pattern reflects the local atomic structure, so does the value of $V(Q, \mathbf{r})$ – a higher $V(Q, \mathbf{r})$ corresponds to a more ordered structure[30]. In this experiment, $V(Q, \mathbf{r})$ was measured from diffractions whose spatial frequency ranged from 3.0 to 5.8 nm$^{-1}$. We measured $V(Q)$ for all scanning points and averaged it over time. Fig. 3(a) shows the calculated fluctuation map at 633 K. The values range from 0.45 to 0.75 and show heterogeneous distribution indicating the coexistence of different local atomic structures. This image directly visualizes the structural heterogeneity of a metallic glass. Thus, it was confirmed that the local atomic structure was heterogeneous. As shown in Figs. 3(b)-(e), although the $V(Q)$ values vary, the structural heterogeneity has been observed at different temperatures.

The temperature dependence of average $V(Q)$ is shown in Fig. 3(f). The average $V(Q)$ slightly decreased with temperature from 0.56 to 0.52 reflecting the decrease in atomic order.

As mentioned above, the dynamical and structural heterogeneities can be measured simultaneously by 5D-STEM. The structure-dynamics relation was evaluated from these heterogeneities by correlating the maps of relaxation time and $V(Q)$. Since the local order was probed by the speckle pattern and the local relaxation time was measured from the temporal change of the same pattern, the local atomic order and atomic rearrangements were measured from the same nanostructure, and the correlation between dynamic and structural heterogeneities directly reflects the structure-dynamic relation. The degree of relaxation depends both on the relaxation time and the value of $\gamma$[31]. To simplify the analysis, considering that $\gamma$ values are close to 1 (see Figs. S1 and S2), we fixed $\gamma$ at 1.0. The correlation was calculated by Spearman's rank correlation, where the correlation value can range from -1 to 1. Positive and negative values mean positive and negative correlations between the maps of relaxation time and $V(Q)$, while 0 means lack of correlation. A scatter plot is shown in Figs. 4 with $V(Q)$ on the vertical axis and relaxation time on the horizontal axis. At 633 K, the distribution of the points is stretched from the center to the upper right, and the correlation value is 0.39, statistically indicating that regions with a long relaxation time and regions with a partial order tend to coincide. Thus, ordered regions tend to have slow dynamics. This tendency is also observed in simulations of polydisperse particles[14]. It is noteworthy that not all structural heterogeneities exhibit correlations with dynamic heterogeneities. Structural heterogeneities calculated by density or potential energy fields do not exhibit a positive correlation with dynamic heterogeneities[14,32]. Simulation studies have demonstrated that the multi-body effect is significant in explaining dynamic heterogeneity, as the relaxation process is affected by surrounding atoms or particles. The diffraction or speckle pattern formed by the interference of electrons scattered by atoms naturally includes information about surrounding atoms[22]. Therefore, a positive correlation has been observed between structural heterogeneity measured by $V(Q)$ and dynamic heterogeneity. The same tendency is observed at higher temperatures as shown in Figs. 4(b)~(e), and especially in Figure 4(f), which superimposes the scatter plots for all temperatures. Figure 4(f) reveals that the relaxation time shortens at higher temperatures. Meanwhile, the shift in the distribution toward the lower left corner indicates that the number of less regular structures that exhibit a higher atomic mobility increases at high temperatures.

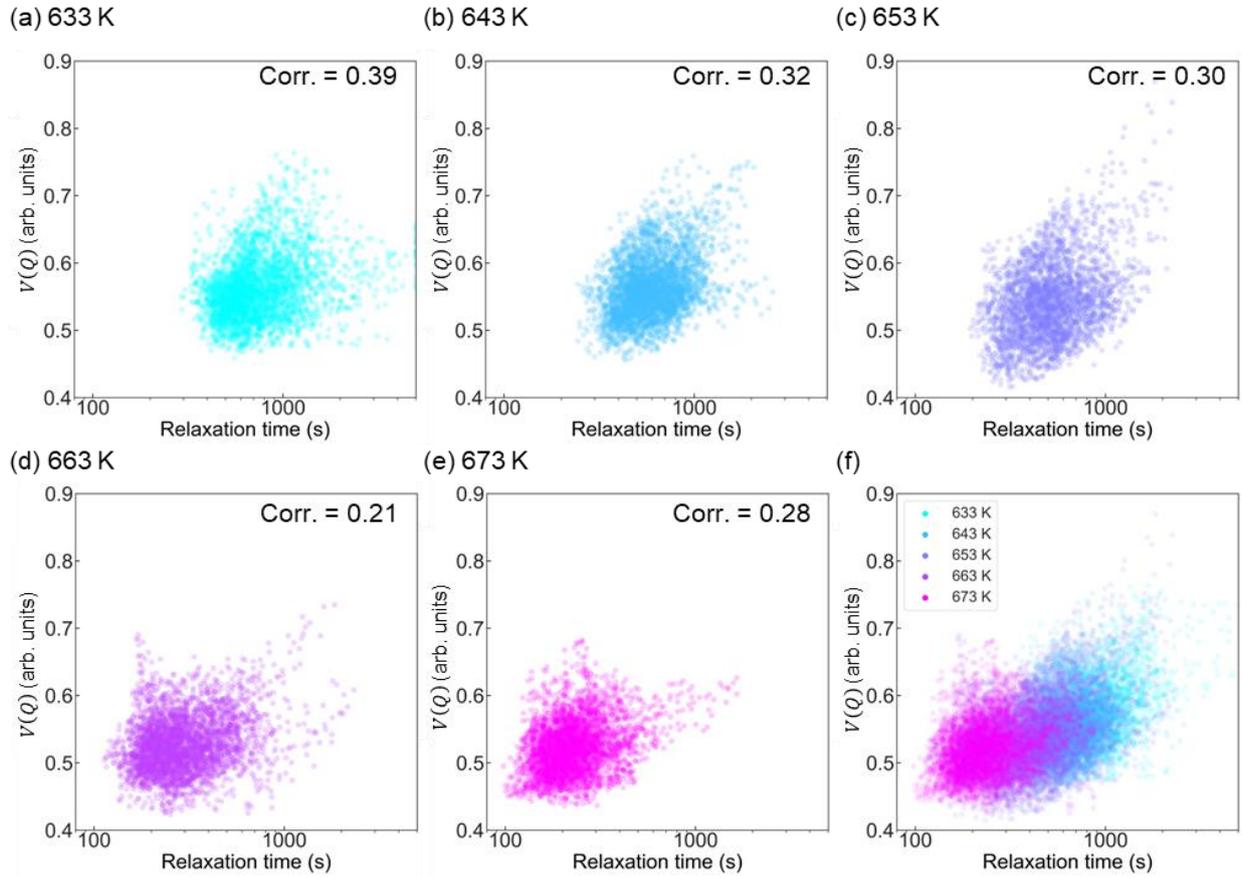

**Fig. 4**. Scatter plots of relaxation time and $V(Q)$ at (a)633K, (b)643K, (c)653 K, (d)663 K, (e)673 K. (f) Superimposed scatter plots at all temperatures.

In summary, our results can be interpreted as follows: at low temperatures, the atomic rearrangement was slow. As the temperature increased, the atomic motion sped up and the degree of local atomic order decreased. The atomic motion was slow for ordered regions regardless of temperature. The atomic motion accelerated near $T_g$ for two reasons: 1) an increase in the number of structures that have a low degree of order and hence a high atomic mobility, and 2) an increase in the atomic mobility in structures with any degree of order.

**Conclusions**

Structural and dynamical heterogeneities are important for elucidating the mechanism of glass transition. In this study, we simultaneously visualized the dynamic and structural heterogeneities in a $Zr_{50}Cu_{40}Al_{10}$ metallic glass by recording the spatiotemporal distribution of diffraction patterns via 5D-STEM. We also measured the temperature dependences of relaxation time and structural order. The results directly reveal a positive correlation between dynamic and structural heterogeneities, indicating that atomic rearrangement is slower in more ordered atomic structures. This study demonstrates the high potential of 5D-STEM in monitoring the dynamic and structural heterogeneities in glasses.

**Materials and Methods**

$Zr_{50}Cu_{40}Al_{10}$ metallic glass was fabricated by the tilt-melting method. Its glass transition and crystallization temperatures were 673 K and 750 K at the heating rate of 1.4 min/K[27]. TEM samples were fabricated by focused ion beam milling.

STEM observation was conducted at 200 kV in an aberration-corrected JEM-ARM200F microscope (JEOL. Ltd.) equipped with a 4DCanvas camera (JEOL. Ltd.). In the normal operation condition, the convergence semi-angle of the electron probe was in the range of 10 to 30 mrad (which corresponds to 6.0 $nm^{-1}$ to 18.0 $nm^{-1}$). This value is much larger than the typical diffraction angle of speckles, and hence it was not possible to observe them. Therefore, the condenser lens and one of the transfer lenses of the corrector were adjusted to achieve an appropriate convergence semi-angle of 1.6 mrad (0.95 $nm^{-1}$) that results in a probe diameter of 0.78 nm (full width at half maximum). We used a beam stop to save the detector from transmitted electron beam. The shadow of beam stop was excluded from the analysis. Each diffraction pattern was acquired within 1 ms at 633 K and 643 K and within 0.5 ms at 653 K, 663 K and 673 K. The number of scan points and the step were 60 × 120 and 0.16 nm, respectively. Total observation area was 9.7 × 19.4 nm. Since 5D-STEM requires a sample edge for data calibration, we set the observation area as a horizontal rectangle. A total of 106 diffraction patterns were acquired for each scan point with a 9.2 s interval at 633 K and 643 K and a 4.8 s interval at 653 K, 663 K and 673 K. The total observation time was 1,048 s at 633 K and 643 K and 524 s at 653 K, 663 K, and 673 K. The pixel size and number of pixels for diffraction patterns were 0.058 $nm^{-1}$ and 264 × 264. The diffraction patterns were 4 × 4 binned to increase the signal to noise ratio. The resultant pixel size and the number of pixels were 0.23 $nm^{-1}$ and 66 × 66.

A fusion heating holder (Protochips Inc.) was used for *in situ* heating experiments. 5D-STEM data were acquired at six temperatures starting from 633 K to 673 K with a 10 K step at a heating rate of 5 K/min. Before the data acquisition, the sample was equilibrated at the target temperature for 1,800 s at 633 K and 643 K, 900 s at 653 K and 663 K and 600 s at 673 K. At each temperature, the equilibration time was at least twice of the average relaxation time.

Sample thickness was measured by electron energy loss spectroscopy (EELS) using the log-ratio method [33] and an inelastic mean free path of 110 nm. The convergence semiangle in this measurement was 20 mrad. The thickness of the observed area ranged from 8 to 50 nm (see Fig. S1).

The diffraction patterns from the left part of the sample were too noisy for reliable measurements. Therefore, we used only the right half of the observed area in the analysis (see Fig. S1).

**Acknowledgement**

We thank Ms. K. Shiomi for preparing the sample and Prof. T. Sannomiya for adjusting the STEM lenses. We also would like to thank Prof. T. Ichitsubo and Prof. H. Kato for helpful discussions.

Supplemental: Structure-Dynamics Correlation in Metallic Glass Revealed by 5-Dimensional Scanning Transmission Electron Microscopy

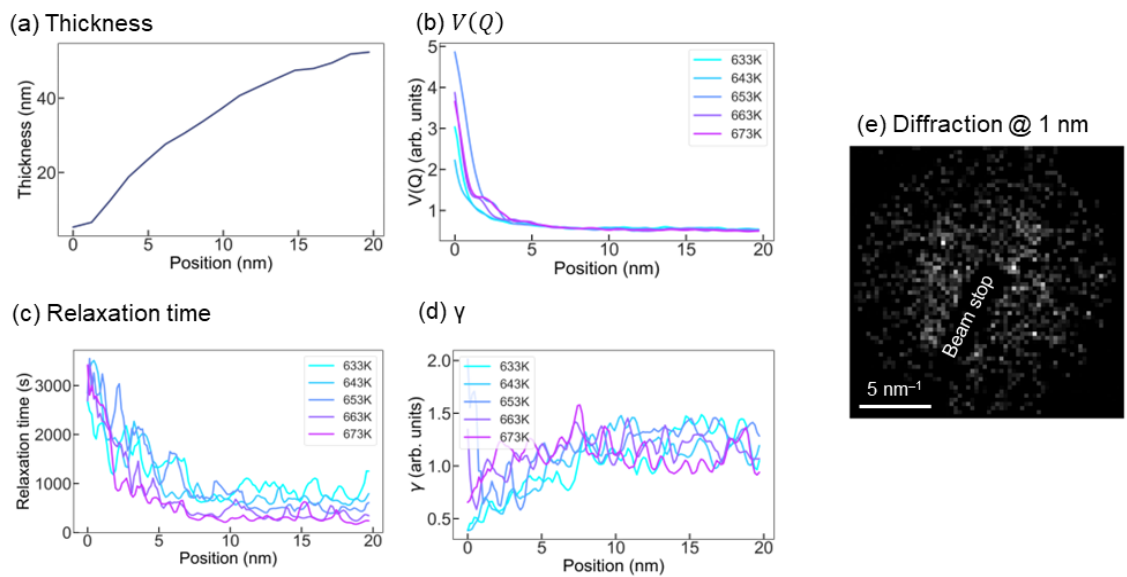

Fig. S1 (a) Thickness, (b) $V(Q)$, (c) relaxation time, and (d) $\gamma$, measured along the spatial x-axis and averaged along the y-axis. As shown fig. S1(e), diffraction patterns acquired within 1 nm from the left edge of the sample exhibit no speckles and are dominated by noise around 4 nm$^{-1}$ because of the small thickness. The values of $V(Q)$, relaxation time, and $\gamma$ measured in this area are not reliable. Strong thickness dependences of $V(Q)$, relaxation time, and $\gamma$ are observed at positions from 1 nm to 10 nm. As the thickness increases, relaxation time and $\gamma$ increase and $V(Q)$ decreases in this sample area. Thickness dependences saturate at positions > 10 nm. Only the area which positions > 10 nm was used for measuring the average relaxation time, $\gamma$, $V(Q)$ and correlation between heterogeneities to reduce the thickness dependence.

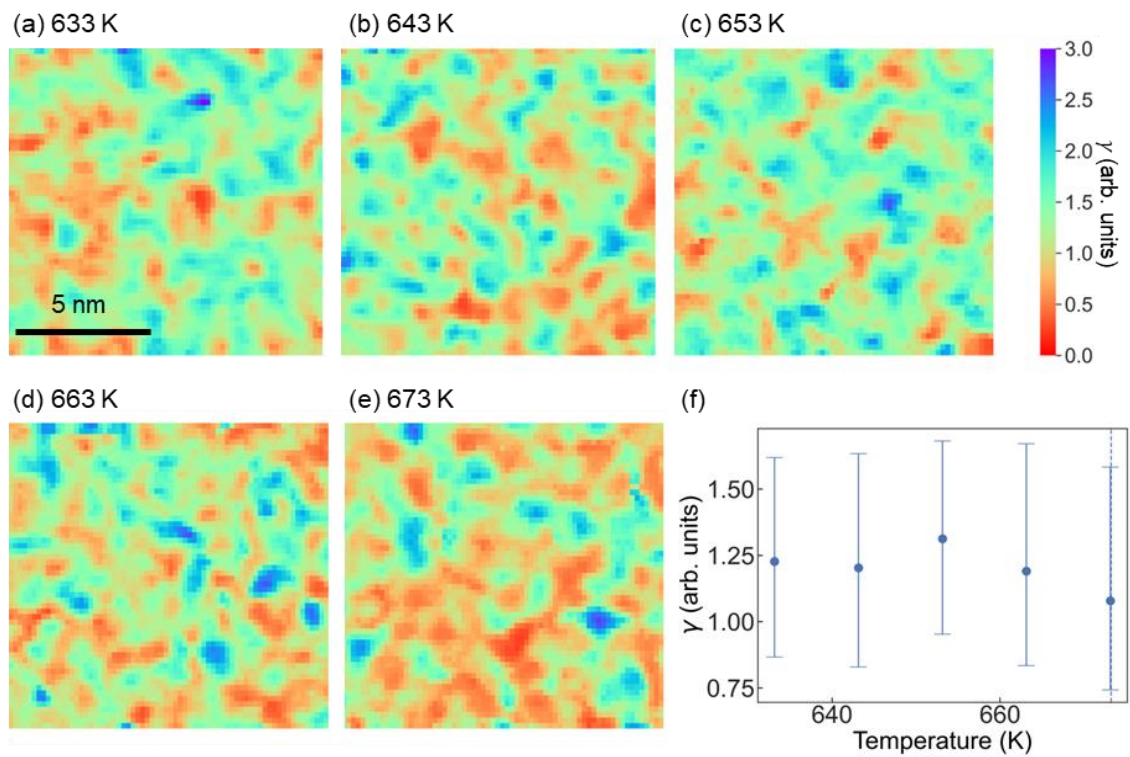

Fig. S2 Maps of $\gamma$ at (a) 633 K, (b) 643 K, (c) 653 K, (d) 663 K, and (e) 673 K. (f) Temperature dependence of the average value of $\gamma$.

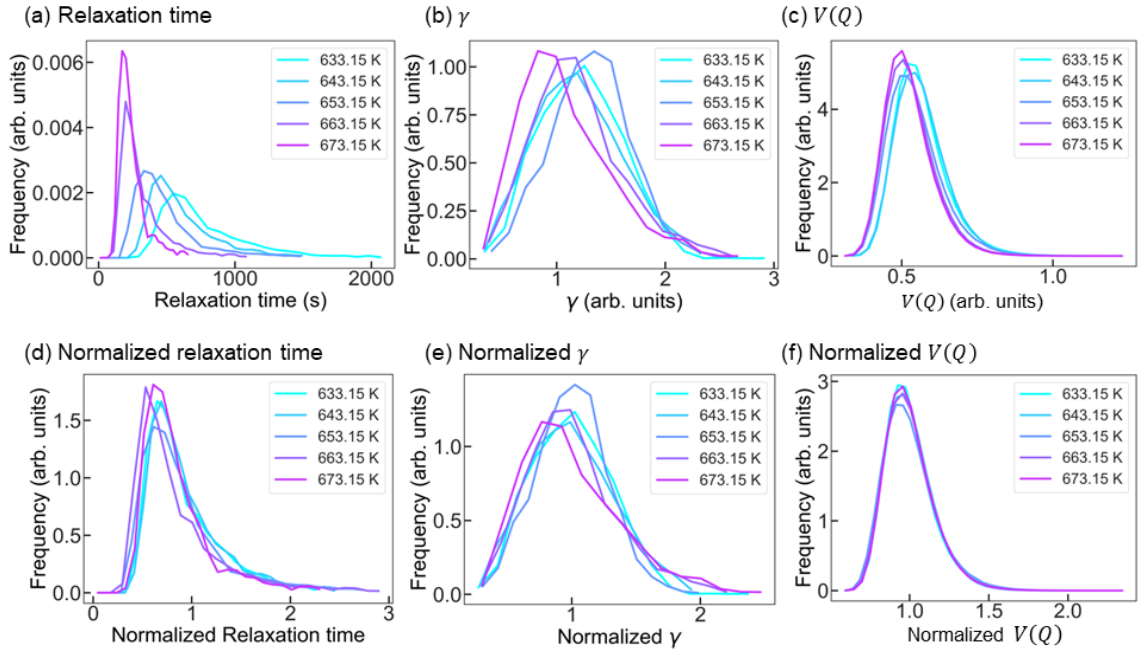

Fig. S3 Distributions of (a) relaxation time, (b) $\gamma$ and (c) $V(Q)$. Panels (d), (e), and (f) show the distributions of normalized relaxation time, $\gamma$, and $V(Q)$. Normalizations are conducted by divisions by their average values.

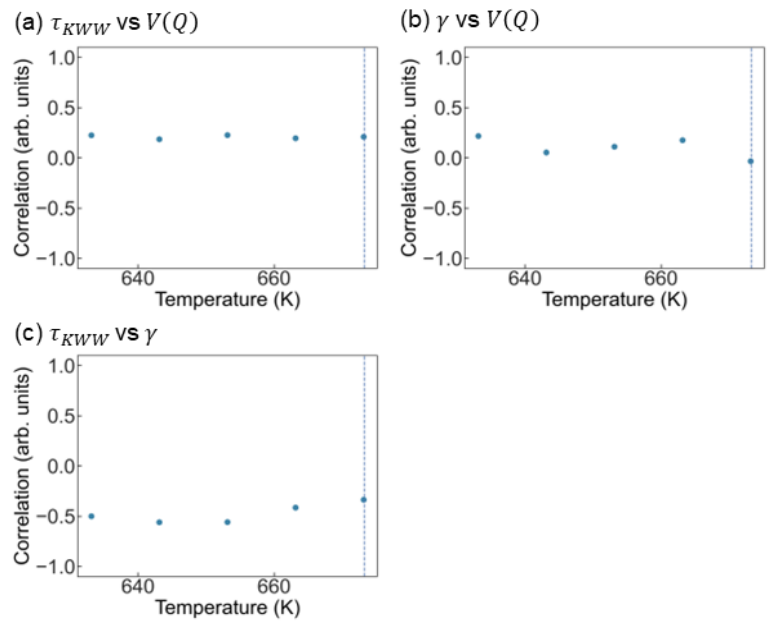

Fig. S4. Temperature dependences of pairwise correlations of (a)$\tau_{KWW}$ vs $V(Q)$, (b)$\gamma$ vs $V(Q)$, and (c)$\tau_{KWW}$ vs $\gamma$.

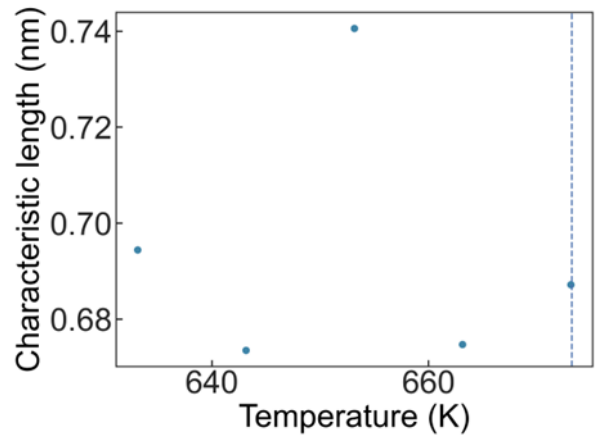

Fig. S5. Temperature dependence of the characteristic length of dynamical heterogeneity.

Measurement of relaxation time.

In calculating the correlation $G(Q, t_1, t_2)$, diffraction patterns were binned by 2 along the time direction to reduce noise. After binning, the number of temporal diffraction images became 53. Gaussian blur with a variance of 0.21 nm$^{-1}$ was applied to diffraction patterns. Then $G(Q, t_1, t_2)$ was calculated from pixels whose $|Q|$ ranged from 4.2 to 4.6 nm$^{-1}$. $g(Q, t)$ was calculated from $G(Q, t_1, t_2)$ and the relaxation time was measured by fitting the KWW function. To reduce noise, $g(Q, 0)$ and $g(Q, t > t_{half})$ was excluded from fitting. $t_{half}$ indicates half of the observation time. We set the range of relaxation time from 0 to 5,000 s and the range of γ from 0 to 3, and performed fitting.

Thickness measurement

Thickness was measured by EELS using the log-ratio method and a convergence semiangle of 20 mrad. We calculated the absolute thickness taking 110 nm as the mean free path of $Zr_{50}Cu_{40}Al_{10}$. This value is the average of mean free path values for each element weighted by its composition[1]. The thickness variation across the sample and thickness dependences of $\tau$, $\gamma$ and $V(Q)$ are shown in fig. S1; $\tau$, $\gamma$ and $V(Q)$ were averaged values over a region of the same thickness. The sample is thicker at the right side than at the left side. The values of $\tau$ and $V(Q)$ at the left edge are excessively high because of the weak signals originating from a thin sample area. In the right half of the sample, $\tau$ and $V(Q)$ increase with thickness. Thickness dependences of $V(Q)$, relaxation time, and $\gamma$ saturate at positions larger than 10 nm from the left edge, indicating a small thickness dependence in the range 35 to 50 nm. To reduce the effect of thickness, only the right half of the sample was used for measuring the average relaxation time, $\gamma$, $V(Q)$, characteristic length and correlation between heterogeneities.

Mapping of γ

Spatial distributions of $\gamma$ are shown in Figs. S2. They are heterogeneous at all temperatures. The temperature dependence of average value of $\gamma$ is shown in Fig. S2(f). Note that $\gamma$ represents the nonlinearity and is expected to be smaller than 1 according to the heterogeneous dynamic scenario. However, at low temperatures, the average of $\gamma$ is around 1.2. As temperature approaches $T_g$, average $\gamma$ approaches 1.0 indicating the decrease of linearity. Thus, in this experiment, the decay of correlation function takes compressed shape. Compressed relaxation functions below $T_g$ is also observed by X-ray photon correlation spectroscopy in the $Mg_{65}Cu_{25}Y_{10}$ glass[2].

Histograms of $\tau$, $\gamma$ and $V(Q)$

Figures S3 show the histograms of $\tau$, $\gamma$ and $V(Q)$. The frequency means the number of counts for each value. The bottom row shows the histograms normalized by their average values. As temperature increases, the frequency of small values of $\tau$, $\gamma$ and $V(Q)$ increases and the histograms become narrower. Normalization eliminated the temperature dependence for relaxation time, $V(Q)$, and possibly for $\gamma$ (considering the large uncertainty in its value). Thus, these values can be scaled by temperature.

Correlation among $\tau$, $\gamma$ and $V(Q)$

We also measured the positional correlation among $\tau$, $\gamma$ and $V(Q)$. The results are shown in Figs. S4. The correlation between $\tau$ and $V(Q)$ at arbitrary $\gamma$ ranges from 0.15 to 0.27 indicating a weaker positive correlation than correlation between $\tau$ and $V(Q)$ measured at $\gamma = 1$. The correlation between $\gamma$ and $V(Q)$ also ranges from 0.17 to 0.0. The correlation between $\tau$ and $V(Q)$ ranges from -0.74 to -0.56 indicating a strong negative correlation between them. Thus, a long relaxation time relates to a small $\gamma$. The small $\gamma$ ($\gamma < 1$) stretches the decay function and effectively lengthen relaxation time and *vice versa*. The correlation between $\tau$ and $V(Q)$ is weaker at arbitrary $\gamma$ than at $\gamma = 1$ because small γ and high $\tau$ correspond to large $V(Q)$.

Characteristic length

To estimate the size of the dynamic heterogeneous region, we also measured the characteristic length of structural heterogeneity $\xi$ by calculating a four-point correlation. We used the following equation:

$$G_4(Q, \Delta r, \Delta t) = \frac{\langle I(Q, r, t)I(Q, r+\Delta r, t)I(Q, r, t+\Delta t)I(Q, r+\Delta r, t+\Delta t)\rangle_{k,r,t}}{\langle I(Q, r, t)\rangle_k \langle I(Q, r+\Delta r, t)\rangle_k \langle I(Q, r, t+\Delta t)\rangle_k \langle I(Q, r+\Delta r, t+\Delta t)\rangle_k}$$

The correlation was measured from the part of diffraction patterns whose spatial frequency ranged from 4.2. to 4.6 nm$^{-1}$. We measured $\xi$ by fitting the following equation

$$G_4(Q, r, t) - 1 = \beta \exp\left(-2\left(\frac{t}{\tau}\right)^{\gamma_t}\right) \exp\left(-2\left(\frac{r}{\xi}\right)^{\gamma_r}\right),$$

where $r$ is the position in real space and $\gamma_t$ and $\gamma_r$ are stretch factors. This fitting yields the length and time to lose the correlation between diffraction patterns. The characteristic length at 633 K was 0.69 nm, which is close to the diameter of the electron probe (0.78 nm). The temperature dependence of characteristic length is shown in Fig. S5. At all temperatures, the characteristic length was smaller than the diameter of the electron probe. Thus, no drastic change in the size of the dynamic heterogeneous region near T$_g$ could be detected.